\documentclass[
	aps,
	pra,
	twocolumn,
	showpacs,
	10pt,
	unsortedaddress,
	superscriptaddress,
]{revtex4-1}
\usepackage{graphicx}
\usepackage[usenames]{color}
\def \thetitle {Effective Josephson dynamics in resonantly driven Bose-Einstein condensates}
\renewcommand{\emph}[1]{\textit{#1}}

\definecolor{extlinkcolor}{rgb}{0.7,0.15,0.15}
\usepackage[
	citecolor=blue,           % I don't like the default green for citations
	urlcolor=extlinkcolor,    % a much more decent color than cyan or magenta.
	colorlinks=true,          % colored links instead of boxes around them
	breaklinks=true,
	pdftitle={\thetitle},
	pdfsubject={\thetitle}
	]{hyperref}

\usepackage{empheq}  % fuer die darstellung von Spaltenvektoren u Matrizen
\usepackage{amssymb}
\usepackage{dsfont}

\newcommand{\bfx}{\mathbf{r}}
\newcommand{\bfy}{\mathbf{r}'}
\newcommand{\ket}[1]{|#1\rangle}

\newcommand{\braket}[1]{\langle#1\rangle}

\newcommand{\spop}[1]{#1}  % decoration for single-particle operators

\begin{document}

\title{\thetitle}

\author{M.~Heimsoth}
\affiliation{Departamento de F\'isica de Materiales, Universidad Complutense de Madrid, E-28040, Madrid, Spain}

\author{D.~Hochstuhl}
\affiliation{Institut f\"ur theoretische Physik und Astrophysik, Christian-Albrechts Universit\"at zu Kiel, Leibnitzstr.\ 15, 24098 Kiel}

\author{C.~E.~Creffield}
\affiliation{Departamento de F\'isica de Materiales, Universidad Complutense de Madrid, E-28040, Madrid, Spain}

\author{L.~D.~Carr}
\affiliation{Department of Physics, Colorado School of Mines, Golden, Colorado 80401, USA}
\affiliation{Physikalisches Institut, Universit\"at Heidelberg, Philosophenweg 12, 69120 Heidelberg, Germany}

\author{F.~Sols}
\affiliation{Departamento de F\'isica de Materiales, Universidad Complutense de Madrid, E-28040, Madrid, Spain}

\date{\today}

% 67.10.Jn  ->  transport processes in quantum fluids
% 67.85.De  ->  dynamic properties of Condensates
% 03.75.Lm  ->  Josephson effect quantum mechanics

\begin{abstract}
We show that the orbital Josephson effect appears in a wide range of driven atomic Bose-Einstein condensed systems, including quantum ratchets, double wells and box potentials. We use three separate numerical methods: Gross-Pitaevskii equation, exact diagonalization of the few-mode problem, and the Multi-Configurational Time-Dependent Hartree for Bosons algorithm. We establish the limits of mean-field and few-mode descriptions, demonstrating that they represent the full many-body dynamics to high accuracy in the weak driving limit. Among other quantum measures, we compute the instantaneous particle current and the occupation of natural orbitals. We explore four separate dynamical regimes, the Rabi limit, chaos, the critical point, and self-trapping; a favorable comparison is found even in the regimes of dynamical instabilities or macroscopic quantum self-trapping. Finally, we present an extension of the $(t,t')$-formalism to general time-periodic equations of motion, which permits a systematic description of 
the long-time dynamics of resonantly driven many-body systems, including those relevant to the orbital Josephson effect.
\end{abstract}

\maketitle

\section{Introduction}
In a bosonic Josephson junction (BJJ) two or a few single-particle states are coherently occupied by a macroscopic number of bosons~\cite{InguscioBose1999}.
These systems combine two qualities that make them interesting for experimental and theoretical studies. First, they are a quantum many-body system. Second, the underlying Hilbert space grows only linearly with the total particle number for the case of two single-particle states. Thus BJJs provide a simple framework to study the role of particle interactions in quantum gases.
Due to the possibility of conveniently controlling the classical limit via the particle number, these systems are particularly well suited for the study of some aspects of the quantum-classical crossover~\cite{MillerSignatures1999,HolthausTowards2001,StrzysKicked2008}.

The Josephson effect in Bose-Einstein condensates (BECs) can exist in a variety of qualitatively different forms. In an {\it external} Josephson junction~\cite{AlbiezDirect2005}, the modes involved refer to localized Wannier functions in a double-well potential.
A more robust realization is provided by the so-called {\it internal} Josephson effect~\cite{HallMeasurements1998,ZiboldClassical2010}, where the states are defined  by the different electronic configurations of the gas atoms.
Dissipation due to the exchange of non-condensed atoms leads to Ohmic damping of the Josephson oscillations~\cite{ZapataJosephson1998}.
Bosonic Josephson junctions can also be used for high precision measurement of physical quantities such as temperature~\cite{GatiNoise2006}, weak forces (gravitational~\cite{CorneyWeak1999} or electroweak~\cite{BarguenoMacroscopic2012}), or chemical potential differences~\cite{KohlerChemical2003}.
Promising scenarios for the creation of macroscopic superposition states~\cite{MicheliMany2003} or spin-squeezed states~\cite{DiazDynamic2012} are based on BJJs.
Furthermore, in a recent experiment at NIST~\cite{WrightDriving2013}, a BJJ in an atom circuit was used to measure rotation.

\begin{figure}[tb]
 \includegraphics[width=0.45\textwidth]{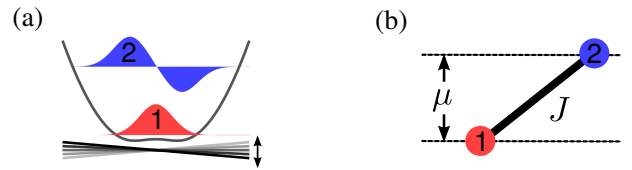}
 \caption{{\it Schematic illustration of an orbital Josephson junction.} A one-dimensional setup is used for illustrative purposes. (a) The driving potential, indicated by the shaded lines below, couples the initial state $\ket{1}$ with an excited state $\ket{2}$. (b) The resulting dynamics is that of a bosonic Josephson junction. In this schematic picture, the horizontal difference represents a chemical potential difference among the modes and the connecting line represents the coupling between them.}
 \label{fig: heuristic scheme}
\end{figure}
Recently, a third kind of Josephson effect in BECs, not classifiable as external or internal, has been identified. In the {\it orbital} Josephson effect (OJE) the single-particle modes have identical internal structure and center-of-mass wave functions with strongly overlapping densities (see Fig.~\ref{fig: heuristic scheme}). The OJE can be realized when a time-periodic driving potential induces a coupling among the unperturbed Floquet states of the system.
Previous work on the OJE~\cite{HeimsothOrbital2012} has focused on the quantum ratchet on a ring potential because the OJE was originally identified in that context. This kind of system is realizable in current BEC setups. In fact an orbital Josephson junction can be obtained in a variety of static and driving potentials.
The present work aims at presenting the OJE in a very general context.
Starting with two very simple illustrative examples, a double well~\cite{AlbiezDirect2005} and a box potential~\cite{GauntBose2013}, we show that the OJE is a general concept that can be studied in a variety of existing experimental BEC systems. Then we explore different driving potentials in ring traps, inspired by recent experimental successes~\cite{ArnoldLarge2006,MunizAxicon2006,MorizotRing2006,RyuObservation2007,HendersonExperimental2009,RamanathanSuperflow2011,MoulderQuantized2012,WrightDriving2013}.

For some selected cases, our theoretical model based on an effective few-mode Hamiltonian is compared to computationally demanding simulations from first principles performed with the Multi-Configurational Time-Dependent Hartree for Bosons (MCTDHB) algorithm~\cite{MasielloMulticonfigurational2005,AlonMulticonfigurational2008,HochstuhlTwo2011}. For the present purposes, MCTDHB can be regarded as a method that enables a full many-body (FMB) study of the interacting boson problem. Thus we focus on the comparison between the truncated description of the orbital Josephson effect and a full many-body calculation that operates in a larger one-atom Hilbert space. Importantly, in both these pictures we work well-beyond the Gross-Pitaevskii (GP) approximation~\cite{LeggetBose2001,pitaevskii2003bose}, which is also considered in some cases.
We find that the \mbox{MCTDHB} results agree well with the truncated BJJ description in a variety of dynamical regimes. This includes regimes that show instabilities in the semiclassical limit such as chaos and unstable fixed points which are poorly described within the GP approximation. Interestingly, we find that the regime of macroscopic quantum self-trapping preserves its character in a non-truncated, full many-body description. This contrasts with the case of an  undriven double-well potential (external Josephson effect), where self-trapping seems to be fragile within a multi-mode picture \cite{SakmannExact2009}.

A major contribution of this paper is the generalization of the so-called $(t,t')$-formalism~\cite{PfeiferAStationary1983,peskin_solution_1993} to an almost arbitrary type of equation of motion including the nonlinear Schr\"odinger equation.
The $(t,t')$-formalism has shown to be an efficient mathematical tool for the study of time-periodic~\cite{peskin_solution_1993,MoiseyevHigh1997} and aperiodic~\cite{MoiseyevAlignment2006} linear systems.
Here we employ the extended $(t,t')$-formalism to analyze the dynamics of field operators in interacting many-body systems.

This paper is arranged as follows.
Section~\ref{sec: generic setup} presents general two-mode orbital Josephson junctions, without specification of the trap geometry or the driving potential. It also includes a rather general presentation of the $(t,t')$-formalism for field operators.
Section~\ref{sec: example systems} introduces two illustrative examples of the orbital Josephson effect: a rocked double well and a box potential perturbed with a modulated lattice.
In Section~\ref{sec: OJE with three modes} the previously introduced concepts are used to explore in greater depth the case of a BEC in a ring trap subject to a ratchet potential. Up to three different driving potentials are discussed, each yielding a three-mode orbital Josephson system.
Finally in Section~\ref{sec: numerical cross check}, we compare the predictions from the effective description in terms of an orbital Josephson junction with full many-body simulations using MCTDHB, focusing on one of the driving potentials introduced in Section~\ref{sec: OJE with three modes}.
In Appendix~\ref{sec: extended tt-formalism} the extended $(t,t')$-formalism is derived in great generality. Appendix~\ref{sec: Convergence study} addresses technical details of the numerical MCTDHB study.

\section{General concepts}
\label{sec: generic setup}
In this section, we present a recipe for the realization of an orbital Josephson junction. We will keep our approach as general as possible.
However, in Section~\ref{sec: OJE with three modes}, we will see that a variety of orbital Josephson junctions also exist which cannot be considered a specific realization of the type presented here.

We consider an initially static BEC, trapped in an arbitrary geometry, given by some static external potential $V_\mathrm{trap}(\bfx)$.
At time $t=0$, a time-periodic driving potential with periodicity $T$,
\begin{equation}
 \label{eqn: potential driving single harmonic}
 V(\bfx,t)=V(\bfx,t+T),
\end{equation}
is switched on.
The dynamics of this many-body system is determined by the Heisenberg equation of motion for bosonic field operators,
\begin{multline}
 \label{eqn: motion Heisenberg general}
 i\partial_t\hat{\psi}(\bfx,t)=\spop{H}_0(\bfx)\hat{\psi}(\bfx,t)
      + V(\bfx,t) \hat{\psi}(\bfx,t) \\
   +\frac{\lambda}{2} \hat{\psi}^\dag(\bfx,t)\hat{\psi}(\bfx,t)\hat{\psi}(\bfx,t),
\end{multline}
where time and spatial coordinates are made dimensionless.
This can be realized by the choice of an appropriate length scale $x_0$ and then expressing all lengths, energies, frequencies, and times in units of $x_0$, $\hbar^2/M_\mathrm{a}x_0^2$, $\hbar/M_\mathrm{a}x_0^2$, and $M_\mathrm{a}x_0^2/\hbar$, respectively, where $M_\mathrm{a}$ is the atom mass.
Furthermore, we set $\hbar=1$, such that energies and frequencies have the same dimensions.
The single-particle Hamiltonian $\spop{H}_0(\bfx)\equiv-\tfrac{1}{2}\nabla^2+V_\mathrm{trap}(\bfx)$ is a sum of kinetic energy and external potential.
The initial field operator $\hat{\psi}(\bfx)\equiv\hat{\psi}(\bfx,0)$ annihilates a bosonic particle at point $\bfx$ and satisfies the standard bosonic commutation relations $[\hat{\psi}(\bfx),\hat{\psi}^\dag(\bfy)]=\delta(\bfx-\bfy)$ and $[\hat{\psi}(\bfx),\hat{\psi}(\bfy)]=0$, which are preserved in time. The contact interaction strength $\lambda$ is proportional to the s-wave scattering length of the gas atoms.

Our goal is to develop an effectively time-independent description that involves only two single-particle states.
We do this by using an extended version of the $(t,t')$-formalism to describe the dynamics of the field operators $\hat{\psi}(\bfx,t)$ in second quantization~\cite{HeimsothOrbital2012}.
The $(t,t')$-formalism has proven to be a powerful tool for solving the Schr\"odinger equation with time-dependent Hamiltonians~\cite{PfeiferAStationary1983,peskin_solution_1993}.
The equation of motion for the field operators~\eqref{eqn: motion Heisenberg general} reads in the extended $(t,t')$-formalism
\begin{equation}
 \label{eqn: motion field operators tt-formalism}
i\partial_t\hat{\psi}(\bfx,t';t)
  =\big[H(\bfx,t')-i\partial_{t'}
   +\lambda \hat{\psi}^\dag\hat{\psi}\big]\hat{\psi}(\bfx,t';t)\,,
\end{equation}
where $t'$ acts as an additional parameter which the field operators depend on.
The single-particle part $H(\bfx,t')\equiv H_0(\bfx)+V(\bfx,t')$ is the sum of kinetic energy, trapping potential, and a driving potential.
Equation~\eqref{eqn: motion field operators tt-formalism}, together with the initial conditions,
\begin{equation}
\label{eqn: initial conditions field opperator tt}
 \hat{\psi}(\bfx,t';0)\equiv\hat{\psi}(\bfx)\text{ for all }\bfx\text{ and }t'
\end{equation}
define a unique solution for $\hat{\psi}(\bfx,t';t)$, from which the physically relevant field operator $\hat{\psi}(\bfx,t)$ can be obtained via
\begin{equation}
 \label{eqn: recover physically meaningfull solution}
 \hat{\psi}(\bfx,t)=\hat{\psi}(\bfx,t';t)\rvert_{t'=t}.
\end{equation}
Note that, the periodicity in $t'$, which is trivially imposed by the initial condition, Eq.~\eqref{eqn: initial conditions field opperator tt}, holds for all times, since the equation of motion~\eqref{eqn: motion field operators tt-formalism} does not break this symmetry.
Furthermore, under these initial conditions the initial commutation relations are also preserved, which read
\begin{align}
  & [\hat{\psi}(\bfx_1,t';t),\hat{\psi}^\dag(\bfx_2,t';t)] = \delta(\bfx_1-\bfx_2),
  \nonumber\\
 \label{eqn: commutation field operators tt-formalism}
  & [\hat{\psi}(\bfx_1,t';t),\hat{\psi}(\bfx_2,t';t)] = 0
  \text{ for all }t'.
\end{align}

We assume the static single particle part $H_0(\bfx)$ to be the predominant term in the equation of motion~\eqref{eqn: motion Heisenberg general}.
This condition is met for systems with weak particle interactions $\lambda$, and a weak overall amplitude of the driving potential~$V(\bfx,t)$.
Furthermore, this assumptions implies, that the low energy eigenstates of the undriven ($t<0$) system are  condensates, with a negligible amount of depletion~\cite{pethick2008bose-einstein} and with the condensate orbitals given by eigenstates of $H_0(\bfx)$.

Within the $(t,t')$-formalism, this means that the predominant part is given by the unperturbed single-particle Floquet operator
\begin{equation}
\label{eqn: Floquet operator single particle}
\mathcal{H}_0\equiv H_0(\bfx)-i\partial_{t'},
\end{equation}
suggesting one ought to use a transformed representation of Eq.~\eqref{eqn: motion field operators tt-formalism} that involves creation and annihilation operators with respect to the unperturbed Floquet states (eigenmodes of $\mathcal{H}_0$).

\subsection{Floquet representation in second-quantized form}
Therefore, we introduce the representation given by
\begin{equation}
 \label{eqn: discrete operator representation}
 \hat{a}_{k,m}(t)
    =\frac{1}{T}\iint d\bfx dt'
        \phi_{k}^*(\bfx)
        e^{i m \omega t'}
        \hat{\psi}(\bfx,t';t),
\end{equation}
where $\phi_k(\bfx)$ denotes an eigenmode of $H_0(\bfx)$.
The corresponding unperturbed single-particle quasienergy is given by
$\mathcal{E}^0_{k m}=\varepsilon_{k}^0-\omega m$, where $\varepsilon^0_k$ is the energy of $\phi_k(\bfx)$.
Quasienergies can be considered the time analog to quasi-momentum from Bloch theory.
The inverse transformation to Eq.~\eqref{eqn: discrete operator representation} reads
\begin{equation}
 \hat{\psi}(\bfx,t',t)
   =
       \sum_{k m}
%        \sum_{k}
%         \sum_{m\in\mathds{Z}}
%         \sum_{m=-\infty}^\infty
        \phi_{k}(\bfx)
        e^{-i m \omega t'}
             \hat{a}_{k m}(t),
\end{equation}
where the sum over $m$ runs over all integer numbers and $k$ indicates the eigenstates $H_0(\bfx)$.
In this representation, the commutation relations~\eqref{eqn: commutation field operators tt-formalism} become
\begin{align}
 \label{eqn: commutation discrete representation tt-formalism}
  &\sum_{m'}\big[
     \hat{a}_{k,m'+m}(t),\hat{a}^{\dag}_{k' m'}(t)
           \big]
  =\delta_{k k'}\delta_{m0}~,\mbox{ and}
   \\
%    \nonumber
  &\sum_{m'}\big[
     \hat{a}_{k,m'-m}(t),\hat{a}_{k' m'}(t)
           \big]
  =0\,,\mbox{for all }m.
\end{align}
In order to obtain an equation of motion for $\hat{a}_{k m}(t)$ we enter these transformations into Eq.~\eqref{eqn: motion field operators tt-formalism}.
For the rest of this section, we focus on driving of the form
\begin{equation}
 \label{eqn: driving for two-level system}
 V(\bfx,t)=\sin(\omega t+\varphi)V(\bfx),
\end{equation}
where the phase $\varphi$ reflects the value of the modulation at the time point $(t=0)$ when the driving potential is switched on.
Driving of this type is particularly relevant for experiments since it can be obtained naturally via inertial forces from shaking~\cite{LignierDynamical2007}, or via intensity modulations~\cite{Salger2009}.
To a first approximation, the latter possibility would cause an additional energy shift of the eigenmodes of $H_0(\bfx)$.

We note that the switching conditions can be tuned to generate the effective gauge factor $e^{i\varphi}$ in Eq.~\eqref{eqn: Hamiltonian general OJJ}.
This generalizes a similar result derived in Ref.~\cite{CreffieldDirected2011} within first quantization.

Note that, in principle, the concepts presented here can be generalized to driving potentials which include higher harmonics.
% Although we will eventually consider higher bichromatic driving, we will focus ...
This yields the equation of motion
\begin{widetext}
 \begin{multline}
 \label{eqn: motion operator discrete representation tt-formalism}
 i\partial_t\hat{a}_{k m}(t)
   =
%    \big( \varepsilon_{k}^0 - \omega m \big)
   \mathcal{E}^0_{k m}
        \hat{a}_{k m}(t)
     +
      \sum_{k' }
        V_{kk' }\frac{1}{2i}
             \big(
               \hat{a}_{k' ,m+1}e^{i\varphi}
               -
               \hat{a}_{k' ,m-1}e^{-i\varphi}
             \big)
   \\
%     +\frac{\lambda}{T}
      +\lambda
       \sum_{(k,m)_{1-3}}
%         \sum_{k_1,k_2,k_3}\sum_{m_1,m_2,m_3}
          \delta_{m+m_1,m_2+m_3}W_{k\,k_1,k_2k_3}\;
          \hat{a}^\dag_{k_1m_1}(t)\hat{a}_{k_2m_2}(t)\hat{a}_{k_3,m_3}(t),
 \end{multline}
\end{widetext}
where the last sum is meant to run over all values of the indices $\{k_1,k_2,k_3\}$, and $\{m_1,m_2,m_3\}$.
It involves the matrix elements of the driving potential,
\begin{equation}
V_{kk' }\equiv \int\!\!\!d\bfx\,\phi_k^*(\bfx)V(\bfx)\phi_{k' }(\bfx),
\end{equation}
 as well as the two-particle matrix elements for contact interaction
\begin{equation}
 \label{eqn: def: two-particle matrix elements}
 W_{k_1k_2,k_3k_4}
  \equiv
    \int\!\!\! d\bfx\;
        \phi_{k_1}^*(\bfx)\phi_{k_2}^*(\bfx)\phi_{k_3}(\bfx)\phi_{k_4}(\bfx).
\end{equation}

We assume the system to be initially condensed in an eigenmode of the undriven single-particle part of the Hamiltonian $H_0(\bfx)$, which will be indicated by the index $k=1$.
This is consistent with the previous assumption of weak particle interactions.
Precisely, the condition $\lambda N\ll \Delta E$ has to be met in order to obtain a sufficiently condensed system; where $\Delta E$ is the energy difference between the initial mode and the neighboring modes.

\subsection{Two-level description in a many-body framework}
In order to obtain a dynamics that is governed by exactly two modes, we impose the following additional conditions on the driving field:
First, there exists a further orbital, indexed as $k=2$, to which the driving potential couples, i.e.\ $V_{1\,2}\neq0$.
Second, the driving frequency $\omega$ is near-resonant, i.e.\ close to the energy difference $\omega_0\equiv \varepsilon^0_1-\varepsilon^0_2$.
This means that the driving frequency can be expressed as $\omega=\omega_0+\Delta$, where $\Delta$ is a possible detuning from exact resonance. It should satisfy $\Delta<\frac{1}{2}|\omega_0-\omega_{1}|$, where $\omega_{1}$ is the resonance closest to $\omega_0$.
Given these two conditions, we say that the driving potential $V(\bfx,t)$ induces a \emph{resonant coupling} between the modes $\ket{1}$ and $\ket{2}$.
As a final assumption, the driving shall not induce any resonant couplings between $\ket{1}$ or $\ket{2}$ and a third state. In practice, this just means any such couplings should be vanishingly small on the time scale of our simulations or the time scales for applying the driving in the case of BEC experiments.
% Finally, the driving $V(\bfx,t)$ shall not induce a coupling between the state $\ket{1}$ ($\ket{2}$) and a third state.

For the case of vanishing particle interactions ($\lambda=0$), it is known that these requirements justify a two-level description~\cite{Grifoni1998229}.
A typical approach to extract the long-time dynamics is given by the rotating wave approximation~\cite{Grifoni1998229}, which, within a single particle description, yields an equivalent, effectively time-independent, two-level Hamiltonian.
In this section, we perform the derivations within the $(t,t')$-formalism instead of using the rotating-wave approximation.
The reason is that in sections~\ref{sec: OJE with three modes} and~\ref{sec: numerical cross check} these concepts will be extended to driving potentials that contain a second harmonic in the time-like part and induce a coupling among three single-particle modes.
Such systems are beyond the standard use of rotating-wave approximation, while the $(t,t')$-formalism can be conveniently applied to more general types of driving.

The truncation to a two-level system can also be justified within the $(t,t')$-formalism.
When the driving frequency is tuned to exact resonance (or sufficiently close to it), the two modes $\ket{1,0}$ and $\ket{2,1}$ become degenerate or nearly degenerate with respect to the predominant Hamiltonian~$\mathcal{H}_0$.
Based on the assumptions above, we truncate the system of equations~\eqref{eqn: motion operator discrete representation tt-formalism}, in such a way that only the operators $\hat{a}_{1 0}$, $\hat{a}_{2 1}$, and their Hermitian adjoint participate.
This yields an effective two-level Hamiltonian describing the many-body dynamics of a resonantly driven Bose system, and gives rise to the orbital Josephson effect.

In general, particle interactions induce a coupling to further modes and could invalidate the two-level description.
Nevertheless, the single-particle part imposes a quasienergy difference between both modes $\ket{1}$ and $\ket{2}$ and further modes, which prevent the system from accessing these further modes.
Therefore, it is reasonable to assume that the two-level description remains valid for weakly interacting particles.

Within this truncated space of resonant levels we will omit the second subindex $m$ in the operators $\hat{a}^{(\dag)}_{10}$ and $\hat{a}^{(\dag)}_{21}$, because its value can be deduced from the first subindex $k$,
% i.e.\ $(k,m)\rightarrow k$.
i.e.\ $(1,0)\rightarrow 1$ and $(2,1)\rightarrow 2$.

With a truncation of the set of modes to $1$ and $2$, the commutation relations~\eqref{eqn: commutation discrete representation tt-formalism} become those of standard bosonic creation and annihilation operators:
\begin{align}
 \label{eqn: commutation discrete truncated}
  & \big[
     \hat{a}_{i}(t),\hat{a}^{\dag}_{j}(t)
    \big]
  =\delta_{ij}~,
   \\ \nonumber
  & \big[
     \hat{a}_{i}(t),\hat{a}_{j}(t)
    \big]
  =0\,,\mbox{with }i,j\in\{1,2\}.
\end{align}

In this truncated picture, Eq.~\eqref{eqn: motion operator discrete representation tt-formalism} becomes
\begin{align}
 \label{eqn: motion operator truncated}
 i\partial_t\hat{a}_{1}
   &=
%    =
    -\tfrac{1}{2}\Delta \hat{a}_{1}
%     -\tfrac{1}{2} i v e^{i\varphi} \hat{a}_{2}
    -\tfrac{1}{2} \vartheta \hat{a}_{2}
%             +\frac{\lambda}{T}
            +\lambda
                \Big(
                   w_{1} \hat{n}_{1}
                   +
                   2w_{12} \hat{n}_{2}
                \Big)\hat{a}_{1},
        \nonumber
        \\
 i\partial_t\hat{a}_{2}
    &=\tfrac{1}{2}\Delta \hat{a}_{2}
%     +\tfrac{1}{2} i v^* e^{-i\varphi} \hat{a}_{1}
    -\tfrac{1}{2} \vartheta^* \hat{a}_{1}
%             +\frac{\lambda}{T}
            +\lambda
                \Big(
                   w_{2}   \hat{n}_{2}
                   +
                   2w_{12} \hat{n}_{1}
                \Big)\hat{a}_{2}\,,
\end{align}
where $\hat{n}_i=\hat{a}^\dag_i\hat{a}_i$, and $\vartheta=i e^{i\varphi} V_{1\,2}$. Furthermore, we have used the abbreviations
$w_{1}\equiv W_{11,11}$, $w_{2}\equiv W_{22,22}$, and $w_{12}\equiv W_{12,12}$.
These are the key interaction terms, corresponding to interactions within mode $\ket{1}$, within mode $\ket{2}$, and between modes $\ket{1}$ and $\ket{2}$.
For a contact interaction, the indices in the two-particle matrix elements are symmetric under a swap of the first two or the last two indices.
For that reason, $w_{12}=w_{21}$ has four equivalent permutations of indices, which gives rise to the prefactor $2$.
Terms like $W_{11,12}$, $W_{22,21}$, or other terms with permuted indices do not appear, because of the first Kronecker-delta in the interaction term in Eq.~\eqref{eqn: motion operator discrete representation tt-formalism} makes these two-particle matrix elements vanish.
% This can be considered a conservation law, referring to the total Floquet band number ($m_1+m_2$) of a two-particle state.

Since, within the truncated picture, standard commutation relations~\eqref{eqn: commutation discrete truncated} are reobtained, the equations of motion~\eqref{eqn: motion operator truncated}
can be regarded as Heisenberg equations of motion coming from the truncated two-level-system Hamiltonian
\begin{multline}
 \label{eqn: Hamiltonian general OJJ}
  \hat{H}_\mathrm{2LS}
  = - \tfrac{1}{2}\Delta
         (\hat{n}_{1}-\hat{n}_{2})
    -  \tfrac{1}{2}(
           \vartheta\hat{a}^\dag_{1}\hat{a}_{2}
         + \vartheta^\ast\hat{a}^\dag_{2}\hat{a}_{1})\\
    + \frac{\lambda}{2} u_{1}  \hat{n}_{1}(\hat{n}_{1}-1)
    + \frac{\lambda}{2} u_{2}  \hat{n}_{2}(\hat{n}_{2}-1) \\
    + \lambda
          w_{12}(\hat{N}^2-\hat{N}),
\end{multline}
with
\begin{equation}
 \label{eqn: def: on-mode interaction strenth}
 u_{j}=w_{j}-2w_{12},
\end{equation}
and $\hat{N}$ is the total particle number operator.
For systems with conserved particle number, the last term in Hamiltonian~\eqref{eqn: Hamiltonian general OJJ} has no effect on the dynamics and can be omitted.
%
% In many-body Hamiltonians we often drop constant terms.
The constant term is only needed for relative comparisons of total energy, e.g.\ for systems with varying number of particles.
When it is omitted, Hamiltonian~\eqref{eqn: Hamiltonian general OJJ} is that of a BJJ, with the peculiarity of a mode-dependent interaction strength.
Note that this few-mode picture does not take into account initial depletion of the condensate. Therefore, we expect this effective description to work well in the limit of high particle numbers and small interaction strength $\lambda$, i.e., the mean-field limit $\lambda N = \mathrm{const.}$, $\lambda \rightarrow 0$, and $N \rightarrow\infty$.

If more than two modes are involved, it is still possible to incorporate the particle interactions in the same manner.
However, in general, the particle interactions will include terms that induce a mixture among the modes.
Furthermore, the mathematical expressions for the interaction term become lengthy when the participation of a third mode is incorporated in such a general manner.
Under certain assumptions on the unperturbed Floquet states, it is still possible to obtain a similarly simple truncated description, cf.\ Section~\ref{sec: OJE with three modes}.

\section{Example systems}
\label{sec: example systems}
\begin{figure}[tb]
  \includegraphics[width=0.45\textwidth]{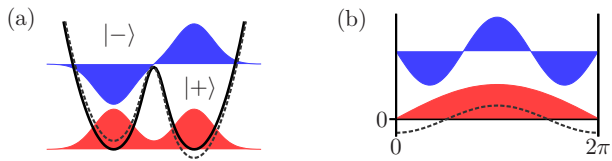}
 \caption{{\it Schematic illustration of the discussed example systems}: (a) driven two-mode system, (b) box-potential with driving. For illustrative purposes, the driven two-mode system (a) is here depicted as a double-well trap. The considerations in Section~\ref{sec: driven two-mode system} may also refer to an internal Josephson system.}
\label{fig: example double well and box potential}
\end{figure}

In this section, we present two examples of specific setups that meet the conditions listed in the previous section and hence represent a realization of an OJE.
A schematic illustration of the systems discussed here is shown in Fig.~\ref{fig: example double well and box potential}.

\subsection{Minimal example: driven two-mode system}
\label{sec: driven two-mode system}

Our first explicit example of an OJE, is set up by a conventional (external or internal) BJJ.
For convenience, we label the modes of this conventional BJJ by $L$ and $R$.
Our starting point is a driven bosonic two-mode Hamiltonian
\begin{multline}
 \label{eqn: Hamiltonian driven LMG}
 \hat{H}(t) =
    -\tfrac{1}{2} J (
             \hat{a}_L^\dag\hat{a}_R
            +
             \hat{a}_R^\dag\hat{a}_L
            )
   + \frac{\lambda}{2}
            \sum_{i=L,R}\hat{n}_i(\hat{n}_i-1) \\
   + \tfrac{1}{2}K
         \sin(\omega t+\varphi)
            (
             \hat{n}_L
            -
             \hat{n}_R
            ),
\end{multline}
where $\hat{a}_i^\dag$ ($\hat{a}_i$) creates (annihilates) a particle in one of the modes $i=L,R$ and obey the usual bosonic commutation relations $[\hat{a}_i,\hat{a}^\dag_j]=\delta_{ij}$.
The Hamiltonian~\eqref{eqn: Hamiltonian driven LMG} can be considered a time-dependent driven version of the Lipkin-Meshkov-Glick model~\cite{lipkin1965validity}.

% The operator $\hat{n}_i\equiv \hat{a}_i^\dag\hat{a}_i$ counts the number of particles found in the mode $i$.
As for the general case discussed above, the driving is switched on at time $t=0$.
The eigenmodes of the undriven single-particle part of Hamiltonian~\eqref{eqn: Hamiltonian driven LMG} are given by the states,
\begin{equation}
 \label{eqn: modes eigen two-site Bose-Hubbard}
 \ket{\pm}\equiv\frac{1}{\sqrt{2}}
          \Big(
           \ket{L}
          \pm
           \ket{R}
          \Big),
\end{equation}
where the signs refer to positive ($+$) and negative ($-$) parity, if they are regarded as localized Wannier states.
Their energy difference is $\varepsilon_{-}-\varepsilon_{+}=J$. Note that $\varepsilon_+<\varepsilon_-$.
For weak driving strengths $K$, and when the system is initially condensed in one of the eigenmodes $\ket{\pm}$, we can apply the recipe presented in Section~\ref{sec: generic setup} and obtain an effectively time-independent description of the system. In this minimal example, the two-particle matrix elements have the simple form
\begin{equation}
 W_{--,--}=W_{-+,-+}=W_{++,++}=\frac{1}{2}.
\end{equation}
The remaining elements of $W$ either can either be obtained by permutation of the indices or are zero.
In any case, the above displayed elements are all we need to apply the prescription~\eqref{eqn: def: on-mode interaction strenth}.
Hence, we obtain an effective inversion of the interaction strength: $u_\pm=-\frac{1}{2}$.
The matrix element of the driving is $V_{+\,-}= \frac{1}{2} K$, from which follows the coupling element $\vartheta= \frac{i}{2} K e^{i\varphi}$.
With $\Delta=\omega-J$, the effective two-level Hamiltonian~\eqref{eqn: Hamiltonian general OJJ} is fully specified.

\begin{figure}[tb]
 \includegraphics[width=0.45\textwidth]{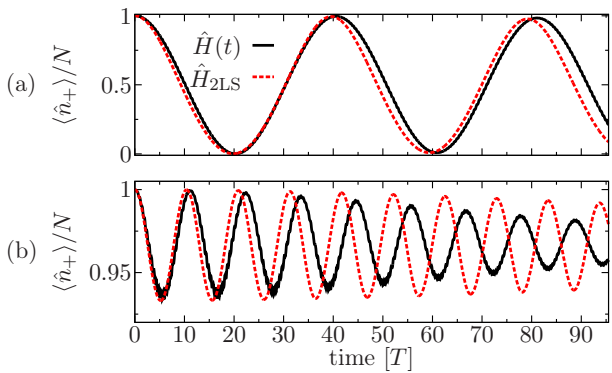}
 \caption{{\it Time-evolution of a driven bosonic two-mode system.} Shown is the normalized occupation of the initial mode $\ket{+}$ in two dynamical regimes, determined by two different mean-field interaction strengths $g\equiv \lambda (N-1)$. (a) Josephson oscillations are observed for weak interaction strength ($g=0.01$). (b) For large interaction strength ($g=0.1$) the system is self-trapped in the initial mode $\ket{+}$ -- note the different scaling on the vertical axes. Parameters are $N=100$, $J=1$, $K=0.05$, $\varphi=0$, $\omega=1$, ($\Delta=0$).}
 \label{fig: example driven doublewell}
\end{figure}

In this minimal example, we already started with a two-mode system.
However, the corresponding effective description as an orbital Josephson system, given by Hamiltonian~\eqref{eqn: Hamiltonian general OJJ}, represents a truncated picture with respect to the unperturbed Floquet states.
% Note that the $(t,t')$-formalism itself is exact.
Figure~\ref{fig: example driven doublewell} shows a comparison of the dynamics governed by the original many-body Hamiltonian~\eqref{eqn: Hamiltonian driven LMG} (dashed curve) and the time-independent effective Hamiltonian~\eqref{eqn: Hamiltonian general OJJ} (solid curve).
The full dynamics features a fast wiggling onset whose frequency is similar to that of the external driving potential.
However, the effective description does not reflect this wiggling. Rather, it refers to the long-term behavior of the system.
Furthermore, there is also a discrepancy in the long-term behavior, which becomes more evident for stronger particle interactions. The precision of this effective description will be discussed in more detail in Section~\ref{sec: numerical cross check}.

\subsection{Box potential}
\label{sec: box potential}
A second simple example is a BEC with a single spatial degree of freedom, trapped in a box potential with length $2\pi$:
\begin{equation}
 \label{eqn: potential box}
 V_\mathrm{box}(x)=
  \begin{cases}
    0,  & \text{for }0\leq x\leq 2\pi
     \\
    \infty, & \text{otherwise}
  \end{cases}.
\end{equation}
% A trap of this type, featuring a flat potential inside an area that is limited by hard-walls has been realized in a recent experiment~\cite{GauntBose2013}.
Such a trap with a flat potential inside the confining walls has been realized in a recent experiment~\cite{GauntBose2013}.

The eigenfunctions of the single-particle Hamiltonian are standing waves
\begin{equation}
 \label{eqn: waves standing}
  \phi_k(x)=\frac{1}{\sqrt{\pi}}\sin(k x/2),
\end{equation}
where $k$ is a positive integer, and the related energy eigenvalue is $\varepsilon_k=k^2/8$.
In this basis, the two-particle matrix elements for contact interaction are given by
\begin{align}
 \label{eqn: two-particle matrix elements box potential}
 w_{k l,k' l' }
  =
    \frac{1}{2\pi}\Big[
    & \delta_{|k-l|,|k' -l' |}
        \big(1+\delta_{kl}\big)
    + \delta_{k+l,k' +l' }
     \\ \nonumber
  - & \delta_{|k-l|,k' +l' }
  -   \delta_{k+l,|k' -l' |}
   \Big].
\end{align}
For the spatial part of the driving potential, we choose the lattice potential
\begin{equation}
 V(x)=\cos(\kappa x/2),
\end{equation}
where $\kappa>0$ is an integer valued parameter.
The matrix elements, with respect to the standing waves~\eqref{eqn: waves standing} read
\begin{equation}
 \label{eqn: matrix elements driving potential box}
  \braket{k|V|l}
      =
       \tfrac{1}{2}
              \delta_{\kappa,|k-l|}\,.
\end{equation}
This means that the potential $V(x)$ induces a coupling among those modes that have momentum difference ($k-l$) of magnitude $|\kappa|$.
A condensate that is initially prepared in a certain condensate state $\ket{k_0}$, will couple either to the mode $\ket{k_0+\kappa}$ or to $\ket{k_0-\kappa}$, if any.
A simultaneous coupling to both modes is ruled out, because the driving frequency $\omega$ cannot match both energy spacings.

At this point, it becomes clear why a harmonic trap is not suited for the realization of an OJE. For a harmonic trap, the spacing of neighboring energy levels is constant.
Consequently, a resonant driving frequency would, in general, induce a subsequent coupling to all eigenmodes of the system.
However, in actual experiments traps are only locally harmonic. Optical traps for example are built on Gaussians~\cite{BlochManyBody2008}, and harmonic plus quartic traps have also been used~\cite{BretinFast2004}.
So in general one does not require a box potential, and a series of other extant experimental systems are viable for the OJE.
The example here can be adapted straightforwardly to those systems.

As in the previous example (Subsection~\ref{sec: driven two-mode system}), for the box potential, the interaction strength is effectively attractive, with $u_{k}=-1$ independent of the specific modes $k$ that form the orbital Josephson system.

\section{Orbital Josephson systems with three modes}
\label{sec: OJE with three modes}
In the following sections, we focus on a BEC in a ring trap, and consider three driving potentials that yield a bosonic Josephson system consisting of three angular momentum modes.
Three illustrative examples are considered, suggesting that a variety of possible driving potentials exists which yield an OJE with this type of trapping potential.
The ring trap~\cite{ArnoldLarge2006,MorizotRing2006,MunizAxicon2006,RyuObservation2007,HendersonExperimental2009,RamanathanSuperflow2011,MoulderQuantized2012,WrightDriving2013} shall be such that the motion of the atoms is effectively frozen along the radial degree of freedom.
Consequently, the dynamics can be modeled as a one-dimensional system of length $2\pi$ with periodic boundary conditions.
This means that we choose the radius $R$ of the ring as our natural length scale ($x_0=R$) and express all times, energies and frequencies accordingly -- cf.\ Section~\ref{sec: generic setup}.
% For a ring trap, loaded with $2-5\times10^{5}$ $^{23}$Na atoms
% (mass $M_\mathrm{Na}=22.99\,\mathrm{u}=3.8\times10^{-26}\mathrm{Kg}$), with major radius $R\simeq10\mu\mathrm{m}$~\cite{RyuObservation2007}, the relevant time scale is $t_0\simeq36\mathrm{ms}$. Accordingly, the energy difference between the zero angular momentum mode and the first nonzero angular momentum mode corresponds to a frequency of about $900\mathrm{Hz}$.
% A driving potential that couples between the initial state and the first excited state would have a resonance at this frequency.

Initially, the potential energy along the single degree of freedom is flat. Therefore, the eigenstates of the single-particle part of the Hamiltonian are plane waves (angular momentum eigenstates),
\begin{equation}
 \label{eqn: plane waves}
  \phi_k=\frac{1}{\sqrt{2\pi}}\exp (i k x),
\end{equation}
characterized by a wave vector $k$ and corresponding energy eigenvalue $\varepsilon_k=\frac{1}{2}k^2$.
Initially, the Bose gas is assumed to be fully condensed in the $k=0$ mode.
At time $t=0$, a weak driving potential is switched on, inducing coupling to modes with $k\neq0$.
% We will call the modes, which the driving potential couples to outer modes.
Optimally, the applied potential provides the possibility to control the coupling strength to each of these modes separately.
We consider three driving potentials, denoted as cases (a), (b) and (c), that combine these qualities.
These are
\begin{widetext}
 \begin{subequations}
 \label{eqn: potential driving ring}
  \begin{align}
 \label{eqn: potential driving ring I}
   &V^a(x,t)=K[\sin(\kappa x)+\alpha\sin(2 \kappa x+\varphi)]
%                  \cdot
              [\sin(\omega t)+\beta\sin(2\omega t+\vartheta)]\,,
   \\
 \label{eqn: potential driving ring II}
  &V^b(x,t)
         = K[
                 \sin(\kappa x)  \sin(\omega t)
          +\gamma\sin(2 \kappa x +\varphi)\sin(2\omega t+\vartheta)]\,,
  \\
 \label{eqn: potential driving ring III}
  \mbox{and }
  &V^c(x,t)
     = K_+\cos(\kappa x-\omega t-\varphi_+)
      +K_-\cos(-\kappa x-\omega t- \varphi_-)\,.
  \end{align}
 \end{subequations}
\end{widetext}
Case~(a) has been realized experimentally on an extended optical lattice~\cite{Salger2009}.
The second case~(b) is the minimal driving potential that leads to the same first- and second-order processes as those of case (a).
Finally, for case~(c) the contribution of second-order processes can be neglected, which makes this driving potential particularly suitable to study the interplay between particle interactions and the truncated picture.
For this case, the underlying processes are analogous to Bragg reflection.
In the adiabatic limit ($\omega\rightarrow0$), each of the contributing waves could be regarded as a conveyor belt, and in classical systems they yield a Brownian surfer~\cite{borromeo1998brownian} if dissipation is added.
A further important motivation of the third potential is that it can be straightforwardly implemented in experiments, because it results from the force of inertia that arises from an orbiting motion of the center of the ring along an ellipse.

Common parameters of the potentials $V^a$, $V^b$, and $V^c$ are the wave vector $\kappa$ of the first harmonic, which controls the angular momentum of the modes to which the driving couples.
The driving frequency $\omega$ should be close to a resonant frequency according to the modes to which the driving couples. For $V^{a}$ and $V^{b}$, the overall amplitude is controlled via the parameter $K$, and for $V^c$ we can define an overall amplitude as $K=\frac{1}{2}(K_++K_-)$.
All three of the driving potentials in Eq.~\ref{eqn: potential driving ring} represent a realization of a ratchet potential.
That is, they all break parity and time inversion symmetry~\cite{Denisov_Periodically_2007,CreffieldCoherent2009}.
For $V^{a}$, both symmetries can be broken separately. The time-inversion symmetry is broken for $\beta\neq0$ and $\vartheta\neq\pi/2,\,3\pi/2$ and parity is broken for $\alpha\neq0$ and $\varphi\neq\pi/2,\,3\pi/2$.
This potential was studied in Ref.~\cite{HeimsothWeakly2010} for the case $\varphi=\vartheta=0$.
% For the other two potentials $V^{b}$ and $V^{c}$, the breaking of both symmetries are controlled simultaneously.
For $V^{b}$, both inversion symmetries hold for $\gamma=0$, while for $\gamma\neq0$, time-inversion is broken with $\vartheta\neq\pi/2,\,3\pi/2$ and parity is broken for $\varphi\neq\pi/2,\,3\pi/2$.
Finally, $V^{c}$ simultaneously breaks both symmetries for $K_+-K_-\neq0$. The phases $\varphi_\pm$ have no control over the symmetry properties of $V^{c}$.

\begin{figure}[tb!]
\includegraphics[width=0.46\textwidth]{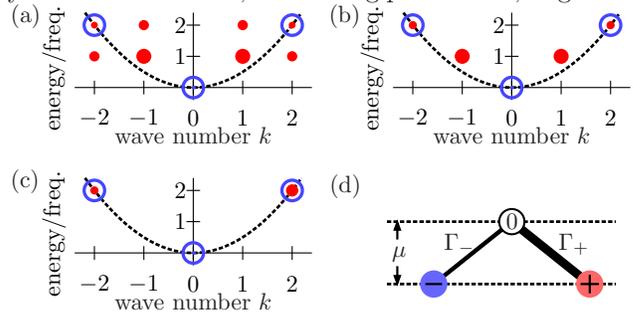}
\caption{
{\it Fourier representation of three driving potentials~\eqref{eqn: matrix elements driving potential ring} that yield an effective description given by Hamiltonian~\eqref{eqn: Hamiltonian effective three level}.}
The label of each panel indicates to which of the driving potentials listed in Eqs.~\eqref{eqn: potential driving ring} it refers to.
The size of the points corresponds of the magnitude of the matrix element $\braket{00|V|km}$.
The lower halves of panels (a-c) are omitted.
Since the potentials~\eqref{eqn: potential driving ring} are real valued, all representation are point symmetric with respect to the origin.
The circles highlight the principal participating modes, i.e.\ those that are {\it intersected} by the parabola of kinetic energy.
The wave vector $\kappa$ of the driving is $\kappa=1$ in (a,b), and $\kappa=2$ in (c).
Panel (d) schematically shows the resulting three-level dynamics.}
 \label{fig: driving potentials Fourier representation}
\end{figure}

All potentials have in common that they induce an OJE involving three modes that can be fully characterized via the angular momentum they carry. These modes are the initial center mode~$\ket{0}$, and two further modes with equal kinetic energy and opposed momenta, denoted as $\ket{\pm}$, cf.\ Fig.~\ref{fig: driving potentials Fourier representation}d.
The coupling strengths for each of the constituent junctions can be tuned separately.
Note that the systems presented in the following cannot be considered a specific realization of the type of orbital Josephson junctions considered in Section~\ref{sec: generic setup}.
They differ in the number of modes that participate in the Josephson system. Furthermore, the driving potential is, in general, not given by a product of a purely spatial and a purely time-like part; and in some cases (a,b), the contribution of a second harmonic in both parts become important.
However, the driving potentials considered here are of course not the only possibilities to obtain an OJE with three modes.
% 
% % 

As in previous sections, we perform our considerations in the $(t,t')$-formalism.
The matrix elements of the driving potentials~\eqref{eqn: potential driving ring} in the discrete representation~\eqref{eqn: discrete operator representation} read
\begin{widetext}
 \begin{subequations}
 \label{eqn: matrix elements driving potential ring}
  \begin{align}
  &V^a_{k m,k' m'}
    = - \frac{K}{4}
       \Big(
          \delta_{k,k' +\kappa}
         -\delta_{k+\kappa,k' }
         +\alpha e^{i \varphi}
             \delta_{k,k' +2\kappa}
         -\alpha e^{-i \varphi}
             \delta_{k+2\kappa,k' }
       \Big)\times
\nonumber
\\&\phantom{SSXXXXXXXXXXXXXXXXXXXXXXX}
       \Big(
          \delta_{m+1,m'}
         -\delta_{m,m'+1}
         +\beta e^{i \vartheta}
             \delta_{m+2,m'}
         -\beta e^{-i \vartheta}
             \delta_{m,m'+2}
       \Big)
      \\
  &V^b_{k m,k' m'}
    =  - \frac{K}{4}
      \Big[
       \big(
       \delta_{k,k' +\kappa}
       -
       \delta_{k+\kappa,k' }
       \big)\big(
       \delta_{m+1,m'}
       -
       \delta_{m,m'+1}
       \big)
      +\gamma
       \big(
       e^{i\varphi}
       \delta_{k,k' +2\kappa}
       -
       e^{-i\varphi}
       \delta_{k+2\kappa,k' }
       \big)\big(
       e^{i\vartheta}
       \delta_{m+2,m'}
       -
       e^{-i\vartheta}
       \delta_{m,m'+2}
       \big)
      \Big]
      \\
  &V^c_{k m,k' m'}
    = K_+\frac{1}{2}
        \Big(
          e^{-i\varphi_+}
          \delta_{k,k' +\kappa}
          \delta_{m,m'+1}
          +
          e^{i\varphi_+}
          \delta_{k+\kappa,k' }
          \delta_{m+1,m'}
        \Big)
    + K_-\frac{1}{2}
        \Big(
          e^{i\varphi_-}
          \delta_{k,k' +\kappa}
          \delta_{m+1,m'}
          +
          e^{-i\varphi_-}
          \delta_{k+\kappa,k' }
          \delta_{m,m'+1}
        \Big).
  \end{align}
 \end{subequations}
\end{widetext}
Figure~\ref{fig: driving potentials Fourier representation} shows a graphical representation of these matrix elements.

For noninteracting particles $V^a$ induces a coupling between the initial mode $\ket{k\,m}=\ket{0\,0}$, and the modes $\ket{\pm2\kappa\,2}$, where the effective coupling parameters can be derived with a perturbative calculation~\cite{HeimsothWeakly2010}.
The second driving potential~$V^b$ is equivalent to $V^a$ within a perturbational approximation that involves processes up to second order in $K$.
Both perturbations $V^{a}$ and $V^{b}$ feature those matrix elements that are involved in the perturbational calculus presented in Ref.~\cite{HeimsothWeakly2010}.
The potential $V^c$ consists of two counter-propagating sinusoidal waves. Each of these waves induces a coupling between the initial mode $\ket{0\,0}$, and $\ket{\pm\!\kappa\,1}$.
In contrast to the potentials $V^{a}$ and $V^{b}$, these couplings can be directly obtained from a first order calculation.

Note that, since the calculations are performed within the $(t,t')$-framework, each mode is characterized by two indices: the quantum number $k$ indicating the angular momentum and the index $m$.
As for the two-level system, within the truncated Hilbert space ($0$, $\pm$) the second number $m$ is determined by the angular momentum and will be omitted.

The dynamics resulting from each of the three drivings in Eq.~\eqref{eqn: potential driving ring} is governed by the effective three-level-system Hamiltonian~\cite{HeimsothOrbital2012}
\begin{align}
 \label{eqn: Hamiltonian effective three level}
  \hat{H}_\mathrm{3LS}
   = &\;\Gamma_+\hat{a}^\dag_+\hat{a}_0
    +\Gamma_-\hat{a}^\dag_-\hat{a}_0
    +\mathrm{h.c.}
     \\
     \nonumber
    &+\mu(\hat{n}_++\hat{n}_-)
    -\frac{\lambda}{4\pi}
                \sum_\nu
                  \hat{n}_\nu(\hat{n}_\nu-1),
\end{align}
given that the total particle number is conserved.
The index $\nu$ in the last sum takes values $\pm$ and $0$.
Here, $\Gamma_\pm$ and $\mu$ are effective parameters that are determined via the parameters of the driving field. Their dependence for each driving potential is listed in Table~\ref{tab: hopping and chemical potential}.
In the case that either $\Gamma_+$ or $\Gamma_-$ vanishes, a two-level system is obtained and the resulting Hamiltonian $\hat{H}_\mathrm{3LS}$ becomes equivalent to Hamiltonian~\eqref{eqn: Hamiltonian general OJJ}, but with a mode-independent interaction term.

\begin{table}[tb]
 \begin{tabular}{lll}
\hline
  & $\Gamma_\pm$  &   $\phantom{-}\mu$   \\\hline
  a\phantom{XXX}  &
  $\frac{K}{4}\big[\frac{K}{2}\pm\alpha\beta e^{-i(\vartheta\mp\varphi)}\big]$ \phantom{XX} &
  $-2\Delta$ \\
  b  &
  $\frac{K}{4}\big[\frac{K}{2}\pm\gamma e^{-i(\vartheta\mp\varphi)}\big]$  &
  $-2\Delta$ \\
  c  &
  $\frac{K_\pm}{2}e^{i\varphi_\pm}$  &
  $-\Delta$ \\
% \hline
 \end{tabular}
\caption{Effective parameters for the truncated picture in dependence of the driving field parameters.
The left column indicates the driving potential $V^{a/b/c}$.
The listed functional dependences refer to the main resonance given by $\omega_0=\frac{1}{2}\kappa^2$ for all potentials and $\Delta$ denotes a possible detuning.
Note that for $V^{a}$ and $V^{b}$ further resonant frequencies exist with a different functional dependence of $\Gamma_\pm$ and $\mu$ on the parameters~\cite{HeimsothWeakly2010}.
}
 \label{tab: hopping and chemical potential}
\end{table}

% The advantage of driving $V^a$ is that the spatial and time-like part factorize. I.e. the time dependence is simply a modulation of the overall amplitude of the spatial part. It has been experimentally realized.

\section{Numerical study of dynamical regimes}
\label{sec: numerical cross check}
An effective description in terms of a BJJ is not guaranteed to always withstand a comparison with a more exact numerical simulation of the full system.
For example, in the case of a BEC in a one-dimensional double well the related two-mode description, referring to localized Wannier functions, has been shown to be invalid near and in the regime where the two-mode model predicts self-trapping~\cite{SakmannExact2009}.
The reasons for that discrepancy are not clear since nominal criteria for the validity of the two-mode description were met in that study.
In the following, we present a numerical check, where some results obtained with the truncated picture of an OJE are compared to the full many-body dynamics.

Because of its rich variety of dynamical regimes, we choose the ratchet system, presented in Section~\ref{sec: OJE with three modes}, for the numerical study.
We focus on the dynamics governed by the driving potential $V^c$ -- see Eq.~\eqref{eqn: potential driving ring III}. This choice rules out possible effects of the intermediate modes that are involved in the higher-order perturbative calculation of the transition amplitudes $\Gamma_\pm$ for the potentials $V^{a}$ and $V^{b}$~\cite{HeimsothWeakly2010}, and allows us to focus on the interplay between interaction and the truncated picture.

We have numerically solved the full many-body (FMB) dynamics involving a large number of modes, using \mbox{MCTDHB}~\cite{HochstuhlTwo2011} for various particle numbers and interaction strengths.
The approximate three-level system~(3LS), determined by Eq.~\eqref{eqn: Hamiltonian effective three level}, is solved numerically via exact diagonalization.
We observe the occurrence of at least three qualitatively different dynamical regimes:
(1)~{\it Rabi oscillations}, between the initial mode $\ket{0}$ and $\ket{a}\propto \Gamma_+ \ket{+} + \Gamma_- \ket{-}$, are expected to occur when the interaction term is weak compared to the driving strength, i.e., $g/4\pi\ll K$, where $g=\lambda(N-1)$ is the mean-field interaction strength.
  The corresponding Rabi frequency is given by $\Omega_\mathrm{R}=2\sqrt{|\Gamma_+|^2 + |\Gamma_-|^2 + \mu^2/4}$. Its inverse $T_\mathrm{R}=2\pi/\Omega_\mathrm{R}$ serves as a natural time scale in the truncated picture.
 (2)~\textit{Chaotic dynamics} are expected for intermediate interaction strength.
 (3)~\textit{Self-trapping} occurs when the particle interactions dominate over the driving strength $g/4\pi\gg K$.
 The critical interaction strength for the occurrence of self-trapping is~\cite{SmerziQuantum1997,HeimsothOrbital2012}
\begin{equation}
 \label{eqn: critical mean-field interaction}
  g_c=8\pi\max(|\Gamma_+|,|\Gamma_-|).
\end{equation}

For our simulations, we choose $\kappa=1$, which allows the use of a small single-particle basis for our FMB simulations (based on \mbox{MCTDHB}).
The single-particle basis, used for the simulation presented here, is given by angular momentum modes, ranging from $k=-4$ to $k=4$.
Further details on the numerical method are presented in Appendix~\ref{sec: Convergence study}.
We choose a small overall driving amplitude $K=0.1$, the regime for which the 3LS approximation should be a good one.
% We choose a driving frequency on resonance
The driving frequency is chosen to be exactly on resonance, i.e.\ $\omega=0.5$ for $\kappa=1$ ($\Delta=0$).
% As driving frequency, we choose the resonant frequency $\omega=0.5$ for $\kappa=1$ ($\Delta=0$).
For the amplitudes of the constituent plane waves of the driving we choose $K_+=0.07$, $K_-=0.03$, and the phases $\varphi_\pm=0$.
We consider four different values of $g$, referring to qualitatively different dynamics.
Those are (a) Rabi oscillations ($g=0.1$), (b) chaos ($g=0.5$), (c) dynamics near the critical point ($g=0.9$), and (c) self-trapped dynamics ($g=1.5$).
The chaotic regime is identified by the calculation of the maximal Lyapunov exponent, which is zero for all other considered cases.
The critical mean-field interaction strength for the self-trapping transition is, according to Eq.~\eqref{eqn: critical mean-field interaction}, $g_c\simeq0.88$.
Particle numbers up to $N=40$ are considered in the following analysis.

For a realistic experimental setup of a ring trap, loaded with $^{23}$Na atoms (mass $M_\mathrm{Na}=22.99\,\mathrm{amu}=3.8\times10^{-26}\mathrm{kg}$), with major radius $R\simeq10\mu\mathrm{m}$~\cite{RyuObservation2007}, the relevant time scale is $t_0\simeq36\mathrm{ms}$. This means that the energy difference between the zero angular momentum mode and the first nonzero angular momentum mode corresponds to a frequency of about $900\mathrm{Hz}$.
A driving potential that couples between the initial state and the first excited state would have a resonance at this frequency.

% An initial condensate depletion due to particle interactions is not taken into account with 3LS.
As was mentioned in Section~\ref{sec: generic setup}, in our 3LS calculations we do not include initial depletion of the condensate.
Within FMB, we can analyze possible effects that originate from the initial depletion.
The FMB simulations, presented in the following, take the ground state of the initial static Hamiltonian as an initial state, including particle interactions.
Consequently, the initial state of the full system is depleted to some extent.
This is a realistic choice for an initial state, since it can be experimentally obtained by the cooling of an interacting Bose gas.
For $N=40$ particles, the initial depletion for the considered interaction strengths stays below $1.4\times10^{-3}$, which is reached for $g=1.5$.
We note that typical BECs in harmonic traps have depletion on this order or smaller~\cite{DalfovoTheory1999}.

\begin{figure}[tb]
 \centering
 \includegraphics[width=0.44\textwidth]{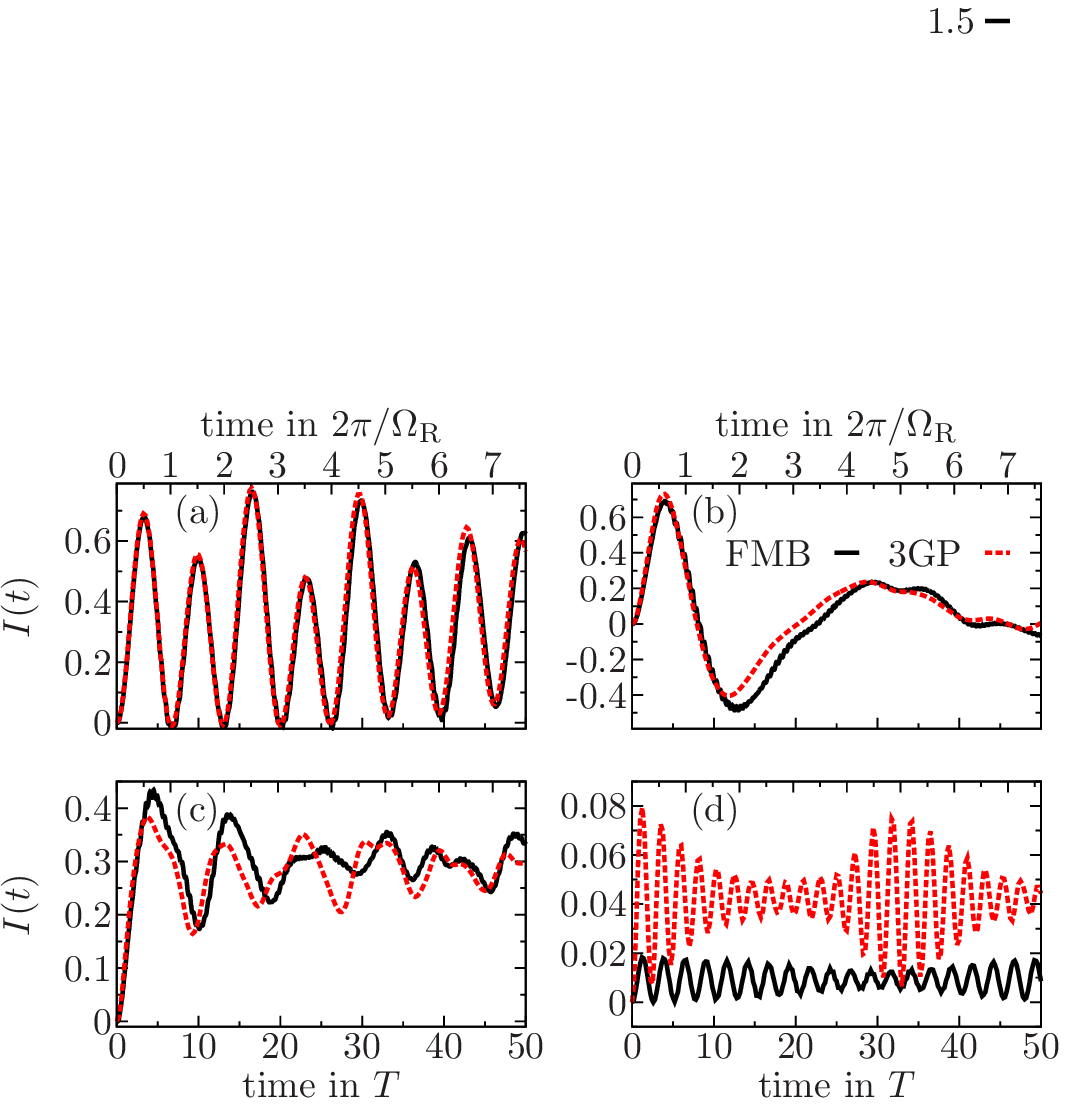}
 \caption{\textit{Instantaneous particle current.}
 Comparison of a full many-body calculation (solid black curve) with the effective three-level description (red dashed curve).
 Four different interaction strengths, which refer to qualitatively different types of dynamics, are considered: (a) Rabi-regime ($g=0.1$), (b) chaos ($g=0.5$), (c) close to critical interaction strength ($g=0.9$), and (d) self-trapped dynamics ($g=1.5$).
 Further parameters are: $N=40$, $K_+=0.07$, $K_-=0.03$, $\varphi_\pm=0$, $\kappa=1$, $\omega=0.5$. }
 \label{fig: comparison of ratchet current}
\end{figure}
\begin{figure}[tb]
\centering
\includegraphics[width=0.47\textwidth]{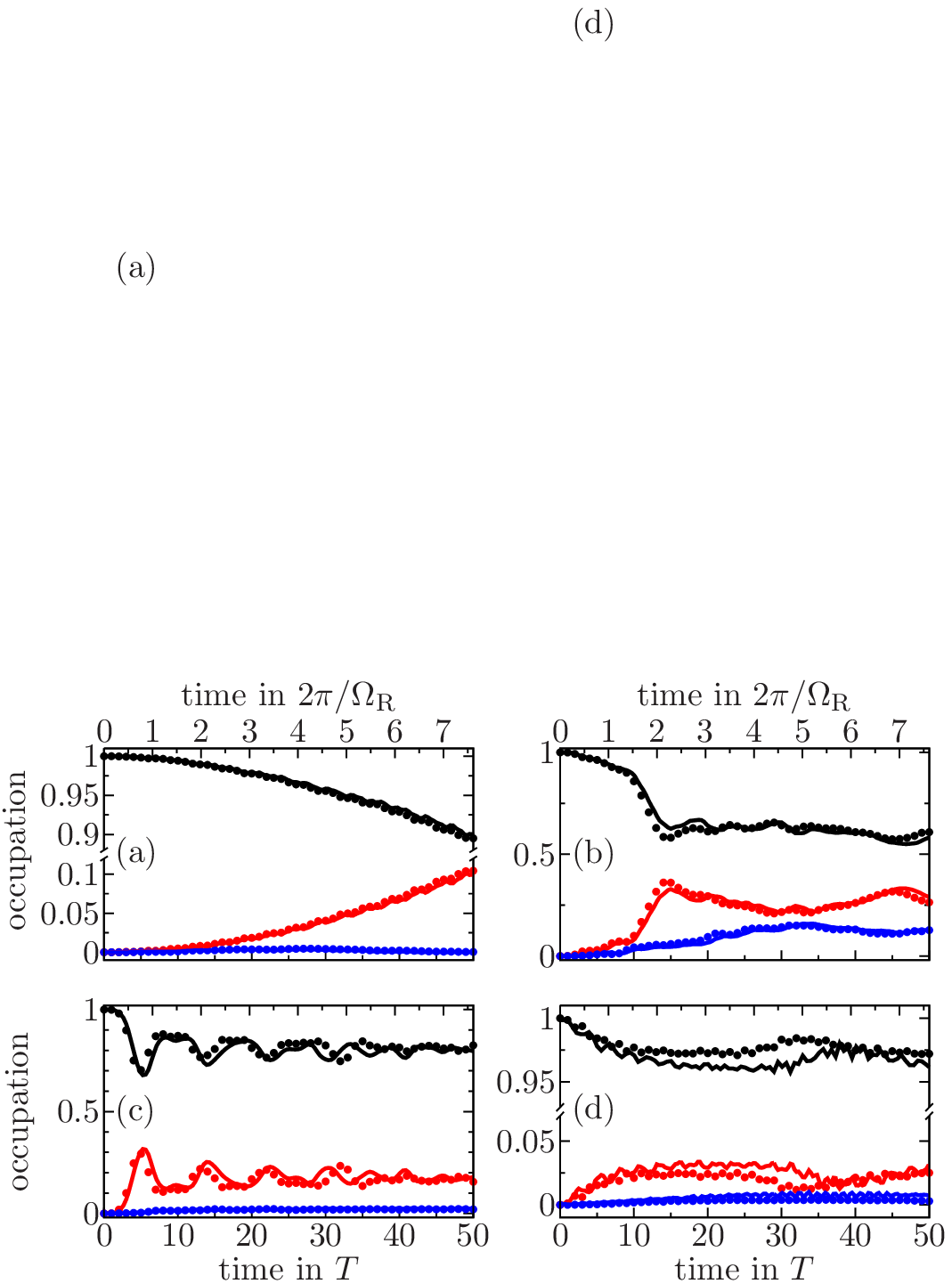}
  \caption{{\it Occupation of natural orbitals.} Solid curves refer to FMB, and the points refer to~3LS.
  The parameters for each panel are the same as in the corresponding panel of Fig.~\ref{fig: comparison of ratchet current}.
  A rapid occupation of more than one natural orbital, as observed in the cases (b) and (c), is an evidence for instabilities within the GP approximation.}
  \label{fig: comparison MCTDHB I}
\end{figure}
\begin{figure}[tb]
\centering
\includegraphics[width=0.47\textwidth]{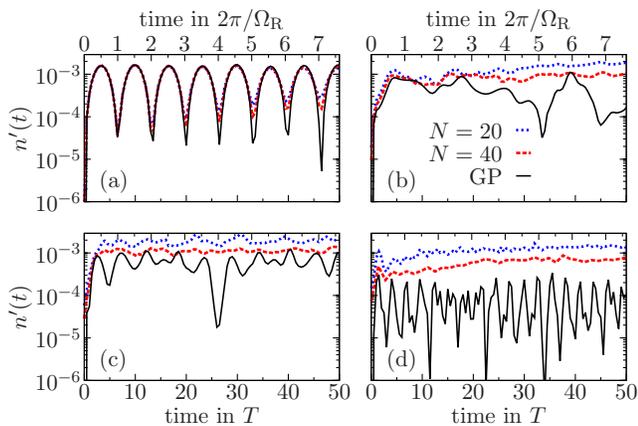}
  \caption{{\it Contribution of outer modes.} Shown is the normalized sum of the occupation numbers of those modes that lie outside the truncated space $(-,0,+)$ -- cf.\ Eq.~\eqref{eqn: def: outer contribution}. Four different mean-field interaction strength are considered (cf. Fig.~\ref{fig: comparison of ratchet current}). The three curves refer to two different total particle numbers and the GP approximation [see key in (b)].}
  \label{fig: MCTDHB outer modes}
\end{figure}
Figures~\ref{fig: comparison of ratchet current} and \ref{fig: comparison MCTDHB I} show a comparison of FMB with 3LS over a time range of $50$ driving cycles, corresponding to more than seven Rabi periods (see upper horizontal scale in Figs.~\ref{fig: comparison of ratchet current} and \ref{fig: comparison MCTDHB I}.).
Figure~\ref{fig: comparison of ratchet current} shows the instantaneous mean current per particle
$I(t)\equiv-i\int dx\braket{\hat{\psi}^\dag\partial_x\hat{\psi}}/N$.
Interestingly, for the unstable cases ($g=0.5$ and $g=0.9$), the discrepancies become quite large at certain times, but both curves revert to a good agreement at $t\simeq35T$.
For example, for $g=0.5$ (Fig.~\ref{fig: comparison of ratchet current}b) the two descriptions deviate by about $40\%$ after $15T$.
The largest discrepancies are observed in the self-trapped regime, where the relative difference
\begin{equation}
\label{eqn: relative difference}
  \sigma(t)\equiv\frac{2|I_\mathrm{3LS}(t)-I_\mathrm{FMB}(t)|}{|I_\mathrm{3LS}(t)+I_\mathrm{FMB}(t)|}
\end{equation}
amounts to $125\%$ at the first maximum ($t\simeq1.2T$). Nevertheless, 3LS correctly reflects the drop of $I(t)$ by at least one order of magnitude as compared to the other cases.

Figure~\ref{fig: comparison MCTDHB I} shows the normalized occupation numbers of the natural orbitals, given by the eigenstates of the single-particle density matrix (SPDM).
The highest occupied orbital is the condensate orbital.
% If more than one orbital is macroscopically occupied, the condensate is said to be fragmented~\cite{SpekkensSpatial1999}.
For a normalized condensate occupation close to $1$, the system can be well described by the GP equation, because this approximation can be derived by making the ansatz of a fully condensed system~\cite{LeggetBose2001,pitaevskii2003bose,alon2007time}.

As can be seen in Fig.~\ref{fig: comparison MCTDHB I}, for $g=0.1$ and $g=1.5$, the system remains condensed during the depicted time range.
For the case of weak particle interactions, this is not surprising, since in the limit of zero particle interactions the system remains fully condensed for all times.
% In both cases, the depletion reaches a maximal value of about 5\%.
For the values of $g$ in between these extreme cases (Fig.~\ref{fig: comparison MCTDHB I}bc), the dynamics is more complex and many-body effects become important.
To be specific, a second orbital becomes macroscopically occupied by an amount of up to 30\%.
This indicates that in these cases, a mean-field treatment would fail and many-body considerations become indispensable.
This observation is consistent with previous works, in which the validity of the GP equation for the case of chaotic dynamics was studied~\cite{CastinInstability1997,BrezinovaWave2011,BrezinovaWave2012,HeimsothOrbital2012}.

One quantification of the relative disagreement between two curves $f_1(t)$ and $f_2(t)$ is given by the time-averaged relative difference
\begin{equation}
\label{eqn: average relative disagreement}
 \bar{\sigma}
   \equiv
  \frac{1}{\mathcal{T}}
  \int_0^{\mathcal{T}}\!\!\!\!d t\,
  \frac{2|f_1(t)-f_2(t)|}{|f_1(t)+f_2(t)|}
\end{equation}
% with $f_1(t)+f_2(t)\neq0$ for $t\in[0,\mathcal{T}]$.
The disagreement, $\bar{\sigma}$, of the condensate occupations, cf. Fig.~\ref{fig: comparison MCTDHB I}, are $0.1\%$, $3\%$, $4\%$, and $1\%$ for $g=0.1$, $0.5$, $0.9$, and $1.5$ respectively.
% These values are larger for the occupations of the less occupied natural orbitals.
% 0.1	0.001  0.11  0.21
% 0.5	0.029  0.15  0.25
% 0.9	0.038  0.24  0.12
% 1.5	0.010  0.37  0.77

Figure~\ref{fig: MCTDHB outer modes} shows the normalized occupation number of the outer modes
\begin{equation}
 \label{eqn: def: outer contribution}
 n'(t)\equiv\frac{1}{N}\sum_{k\notin \{\pm,0\}}\braket{\hat{n}_k}(t),
\end{equation}
i.e., the total occupation those modes that lie outside the three-level Hilbert space.
Two different particle numbers ($N=20$ and $N=40$) and the mean-field approximation are considered for the same values of $g$ as in Figs.~\ref{fig: comparison of ratchet current} and~\ref{fig: comparison MCTDHB I}.
% The relative participation of outer modes does not vary strongly with the total particle number.
A general trend is that the participation of outer modes decreases for larger particle numbers.
Furthermore, we see that the GP equation can also provide some information about the relative participation of outer modes.
A clear trend of an increase of the participation of outer modes in time cannot be distinguished for the depicted cases.
It seems to reach a saturated value after less than one Rabi period with a fluctuation onset, except for the Rabi regime, where the occupation of outer modes shows an oscillation.
In the self-trapping regime, the GP equation clearly underestimates the contribution of outer modes to the dynamics.
The participation $n'(t)$ remains below $3\times10^{-3}$ for all cases depicted in Fig.~\ref{fig: MCTDHB outer modes}.

We have performed additional FMB simulations, with a fully condensed initial state in the $k=0$ mode, in order to see whether some of the discrepancies observed here between 3LS and the full dynamics can be assigned to the initial depletion, present in the initial state of the full system.
For all interaction strengths considered, the discrepancies between the effective 3LS description and the full dynamics exceed the discrepancies observed here between the two versions of the FMB simulations (with and without initial depletion).
Precisely, the relative disagreement between the condensate occupations of the initially depleted and the initially fully condensed system is $\bar{\sigma}=3\times10^{-4}$ for $g=0.1$ and is below $0.7\%$ for the other three considered interaction strengths.
These results suggests that initial depletion of the condensate plays a minor role in the comparison presented here.
% average disagreement between natural occupation numbers for initially depleted and initially fully condensed system
% 0.1	0	0.000349511
% 0.1	1	0.0607099
% 0.1	2	0.0861951
% 0.5	0	0.00643137
% 0.5	1	0.0632933
% 0.5	2	0.0895997
% 0.9	0	0.00428008
% 0.9	1	0.0684718
% 0.9	2	0.195813
% 1.5	0	0.0046968
% 1.5	1	0.152061
% 1.5	2	0.376913

\begin{figure}[tb]
\centering
\includegraphics[width=0.47\textwidth]{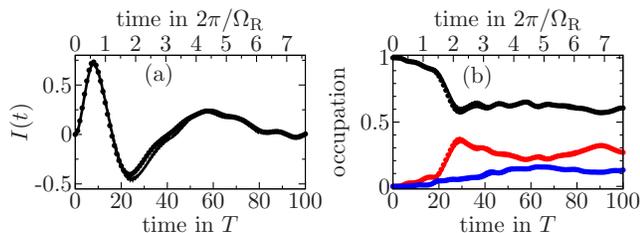}
\caption{{\it Comparison between 3LS and FMB results with a weaker driving amplitude.} Shown is (a) the time evolution of the instantaneous particle current and (b) the time evolution of the normalized occupation numbers of the natural orbitals.
Points refer to 3LS and solid lines show FMB results.
Parameters: $K_+=0.035$, $K_-=0.015$, $g=0.025$, $N=40$. Other parameters are the same as in Fig.~\ref{fig: comparison of ratchet current}.}
 \label{fig: MCTDHB vs 3LS halved K}
\end{figure}

Figure~\ref{fig: MCTDHB vs 3LS halved K} shows a comparison between 3LS and FMB with a weaker driving (halved values of $K_\pm$ or $K$).
The ratchet current is given in Fig.~\ref{fig: MCTDHB vs 3LS halved K}a, and Fig.~\ref{fig: MCTDHB vs 3LS halved K}b compares the occupation of the natural orbitals.
The interaction strength is $g=0.25$, such that the dynamics is completely analogous to the chaotic case shown in Figs~\ref{fig: comparison MCTDHB I}b, but with a larger effective time scale.
Note that the time range in Fig.~\ref{fig: MCTDHB vs 3LS halved K} has doubled, compared to Fig.~\ref{fig: comparison MCTDHB I}.
It shows that both descriptions approach each other for smaller values of the driving amplitude.
For example, the difference in the ratchet current between 3LS and FMB after three Rabi-periods amounts $0.1$ for $K=0.1$, while for $K=0.05$ it has decreased to $0.05$.
Similarly, the disagreement, $\bar{\sigma}$, of the normalized condensate occupation has halved to $1.5\%$.
Thus the 3LS improves in accuracy for weaker driving amplitude.
%\sigma= 0.015 0.092 0.23 (these numbers show that this linear behavior does not hold for all quantities)

\section{Conclusions}
We have shown that the orbital Josephson effect is a general concept that can be realized in a variety of driven condensate setups. The OJE manifests itself when single-particle states occupying the same region of space are coherently populated by a macroscopic number of resonantly driven bosons. It is distinct from the external or internal Josephson effect, which do not require external driving. We have listed the main characteristics that must be met by the trap and driving potentials to realize the OJE. We have discussed several trap geometries and for one of them we have considered three different driving potentials.
Realistic experimental parameters and realizations were provided for these geometries and cases.

In two selected cases, the truncated Josephson description has been compared with the full many-body dynamics that encompasses a larger one-atom Hilbert space.
In some cases, the effective, few-mode description is good approximation of the systems dynamics, and becomes increasingly better for weaker driving amplitudes.
Even in cases where the discrepancies are large, the Josephson description still serves as an efficient way to capture the qualitatively different dynamical regimes and to predict in which parameter regimes they may be found.
For some parameter regimes, many-body considerations (beyond mean-field) are necessary, even if depletion can be initially neglected. We have found that the effective (truncated) description can be correct in those cases as well. A few-mode description remains valid even near instabilities as long as the driving is sufficiently weak. Finally, we would like to remark that the regime of macroscopic quantum self-trapping is not an artifact of the effective description but is preserved within a full many-body calculation over simulation time scales.

We have extended the $(t,t')$-formalism to a rather general case, which permits a convenient description of the coarse-grained, long-time dynamics of resonantly driven many-body systems. The extension can be applied, e.g., to the Heisenberg equation of motion of field operators in an arbitrary representation as well as to the time-dependent nonlinear Schr\"odinger equation.

\acknowledgments
The authors acknowledge support from Spain's MINECO through Grant No.~FIS2010-21372 and the Ram\'on y Cajal program (CEC), the Comunidad de Madrid through Grant Microseres, the Heidelberg Center for Quantum Dynamics (LDC), the Alexander von Humboldt Foundation (LDC), and the U.S. National Science Foundation under grant PHY-1067973 (LDC).

\appendix
\section{Generalization of the (t,t')-formalism to arbitrary equations of motion}
\label{sec: extended tt-formalism}
In reference~\cite{HeimsothOrbital2012} it was shown, how the $(t,t')$-formalism can be extended to the Heisenberg equation of motion for field operators. This extension was nontrivial in many regards, since for interacting many-body systems, the resulting equations of motion are nonlinear in the fields.
In the following, we will show, how this extension can be considered a special case of a more general extension of the $(t,t')$-formalism to any nonlinear system -- Hamiltonian or not.
In previous works, where the $(t,t')$-formalism has been derived, the linearity of the underlying equation of motion was assumed and used.
To be precise, the solution of a Schr\"odinger equation with a static Hamiltonian was expressed via a
unitary time-evolution operator, given by the exponential of the Hamiltonian. Such an approach cannot be used for an extension to arbitrary equations of motion, which is why we will follow a different path here.
We end the appendix by discussing the application of the formalism to the solution of the time-dependent GP equation.

\subsection{General case}
Our goal is to map systems with an underlying time-periodic equation of motion, with period $T$, to systems without explicit time dependence. The dynamics of a physical system is given by a trajectory in the phase space $\mathcal{P}$ of the system.
For now, we do not need to make any further assumptions on $\mathcal{P}$.
It can have finite or infinite dimensions, and does not have to be a Hilbert space, i.e.\ no scalar product or norm needs to be defined on $\mathcal{P}$.
Further below, we will pay special attention to the case of square-integrable wave functions.

The state of the system is given by a vector $v\in\mathcal{P}$.
Its dynamics is fixed by the initial value problem:
\begin{equation}
 \label{eqn: initial value problem general}
 \frac{d}{d t}v(t)=F\big(v(t),t\big),\mbox{ and } v(0)=v_0.
\end{equation}
As mentioned above, we restrict our considerations to time-periodic systems:
\begin{equation}
 \label{eqn: boundary conditions of F}
F(\cdot,t+T)=F(\cdot,t).
\end{equation}
In any other sense, $F$ is arbitrary. It can be nonlinear, discontinuous in $v$ or in $t$, and of course it is allowed to have no explicit time dependence at all, in which case the period $T$ can be chosen freely.

Now we consider an arbitrary {\it generalized loop} in $\mathcal{P}$, defined as
\begin{equation}
 \label{eqn: generalized loop in phase space}
\bar{v}(t') \in \mathcal{P}\mbox{ for all } t'\in[0,T]
\mbox{ and } \bar{v}(0)=\bar{v}(T).
\end{equation}
We allow $\bar{v}(t')$ to be contracted, e.g., to a point ($\bar{v}(t')\equiv\bar{v}_0$) or to a line.
Hence, $\bar{v}(t')$, strictly speaking, does not necessarily form a loop. This is why we assign the term \textit{generalized loop}. This generalized loop may evolve in time, according to the initial value problem
\begin{multline}
 \label{eqn: equation of motion two times}
 \frac{\partial}{\partial t}\bar{v}(t',t)
   = \mathcal{F}\big(\bar{v}(t',t),t'\big)
   \equiv F\big(\bar{v}(t',t),t'\big)-\frac{\partial}{\partial t'}\bar{v}(t',t),
    \\
  \mbox{and}\quad \bar{v}(t',0)\equiv v_0\mbox{ for all }t'.
\end{multline}
Note that $\mathcal{F}$ is a more general object than $F$, since it contains a partial derivative of $\bar{v}$ with respect to $t'$.
The dynamics of this generalized loop describes a surface with the structure of a \textit{generalized tube} in the phase space of the considered system.
Having our goal in mind, We are looking for an equation of motion that determines a parametrized family of curves through this surface. This family, with parameter $\tau$, is given by
\begin{subequations}
\label{eqn: family of curves}
\begin{align}
\label{eqn: recipe for recovery of physical state}
 v'_\tau(t)&\equiv \bar{v}(t'_\tau(t),t),
 \\
 \label{eqn: general cut}
 \text{ with }
 t'_\tau(t)&\equiv (t+\tau)\,\mathrm{mod}\,  T~.
\end{align}
\end{subequations}
Since $\bar{v}(t',t)$ and $\mathcal{F}$ are periodic in $t'$, it is sufficient to restrict ourselves to $\tau\in[0,T]$.
An equation of motion for $v'_\tau(t)$ can be obtained by deriving both sides of Eq.~\eqref{eqn: recipe for recovery of physical state} with respect to $t$.
The first argument of $\bar{v}$ on the right hand side of the definition~\eqref{eqn: family of curves} is itself a function of $t$. Therefore, we have to derive partially with respect to both arguments of $\bar{v}$ and apply the chain rule for the case of the first argument. This yields
% \begin{align}
% \nonumber
%  \frac{d}{d t}v'_\tau(t)
%  &= \frac{d}{d t}\bar{v}(t'_\tau(t),t)
%   = \underbrace{\frac{\partial t'}{\partial t}}_{=1}\frac{\partial \bar{v}(t',t)}{\partial t'}
%    + \frac{\partial \bar{v}(t',t)}{\partial t}\Bigg\rvert_{t'=t'_\tau(t)}
%   \\
%  &=F\big[\bar{v}(t'_\tau,t),t'_\tau\big].
% \end{align}
\begin{align}
 \frac{d}{d t}v'_\tau(t)
  &= \frac{d}{d t}\bar{v}(t'_\tau(t),t) \nonumber\\
  &= \left[\frac{\partial t'}{\partial t} \frac{\partial \bar{v}(t',t)}{\partial t'}
   + \frac{\partial \bar{v}(t',t)}{\partial t}\right]_{t'=t'_\tau(t)}
   \nonumber\\
  &= \left[\frac{\partial \bar{v}(t',t)}{\partial t'}
     +F\big[\bar{v}(t',t),t'\big]-\frac{\partial \bar{v}(t',t)}{\partial t'}\right]_{t'=t'_\tau(t)}
   \nonumber\\
  &=F\big[\bar{v}(t'_\tau,t),t'_\tau\big].
\end{align}
where, $t'_\tau$ is always meant to be a function of $t$, although it is not explicitly expressed.
We have used Eq.~\eqref{eqn: equation of motion two times} to express $\partial \bar{v}(t',t)/\partial t$ as well as the identity $\partial t'_\tau(t)/\partial t=1$.
The modulo operation within the argument of $F(v,\cdot)$ does not have any effect due to the periodicity of $F(v,\cdot)$.
Hence, we obtain the equation of motion for $v'_\tau(t)$, given by
\begin{equation}
 \label{eqn: single time equation of motion on the cut}
 \frac{d}{d t}v'_\tau(t)=F\big(v'_\tau(t),t+\tau\big).
\end{equation}
The initial conditions follow from Eq.~\eqref{eqn: equation of motion two times}: $v'_\tau(0)=v_0$.
This means that for each $\tau\in[0,T]$, $v'_\tau(t)$ is the solution to the initial value problem
\begin{equation}
\label{eqn: initial value problem parametrized.}
 \frac{d}{d t}v(t)=F\big(v(t),t+\tau\big),\mbox{ and } v(0)=v_0.
\end{equation}
In particular, the solution to the initial value problem, given in Eq.~\eqref{eqn: initial value problem general}, is obtained by
\begin{equation}
 \label{eqn: recipe to obtain relevant solution}
 v(t)=\bar{v}(t,t).
\end{equation}

Note that the $t'$-dependence of $\bar{v}$ is originally restricted to the interval $[0,T]$, but as we imposed periodic boundary conditions in Eq.~\eqref{eqn: generalized loop in phase space}, we can directly extend $\bar{v}(t')$ to be defined on the entire real axis.

The extended $(t,t')$-formalism, derived here, is quite general.
The underlying equation of motion could be a linear~\cite{PfeiferAStationary1983,peskin_solution_1993} or a nonlinear differential equation; it could be the Heisenberg equation of motion for field operators~\cite{HeimsothOrbital2012}, the nonlinear Schr\"odinger equation (see Section~\ref{sec: NSE}), or any other equation of motion from a very different context, including systems with dissipation.
% It can be applied to study, e.g., the dynamics of a driven pendulum, with and without dissipation; classical or quantum ratchet systems; or the dynamics of classical fluids under time-periodic forces.
% However, since this thesis focuses on the dynamics of ultracold atomic gases, the discussions in the subsequent sections are restricted to its application to equations of motion that describe quantum mechanical systems.

\subsection{Nonlinear Schr\"odinger equation}
\label{sec: NSE}
As mentioned above, these concepts can be applied to the dynamics of square integrable wave functions. In the case that the underlying equation of motion is linear, the standard $(t,t')$-formalism~\cite{peskin_solution_1993} is obtained. Here, we will focus on the nonlinear Schr\"odinger equation,
\begin{equation}
 \label{eqn: Gross-Pitaevskii}
 i\partial_t\psi(\bfx,t)
  = H_0(\bfx,t)\psi(\bfx,t)+g|\psi(\bfx,t)|^2\psi(\bfx,t),
\end{equation}
where $H_0(\bfx,t)=V(\bfx,t)-\tfrac{1}{2}\nabla^2$ is the operator associated with the linear part of this equation.
We consider it to be time periodic, with period $T$: $H_0(\bfx,t+T)=H_0(\bfx,t)$.
The wave function $\psi$ shall be normalized as $\int\!\!d \bfx \,|\psi(\bfx)|^2 = 1$, which can be shown to be conserved under equation of motion~\eqref{eqn: Gross-Pitaevskii}.

In this case the phase space is the Hilbert space of square integrable complex-valued wave-functions $\psi(x)$.
Applying the recipe as outlined above in Eqs.~\ref{eqn: equation of motion two times}-\ref{eqn: initial value problem general}, we obtain the nonlinear Schr\"odinger equation within the $(t,t')$-formalism:
\begin{multline}
 \label{eqn: Gross-Pitaevskii with two times}
 i\partial_t\psi(\bfx,t';t)
  = \big(
     H_0(\bfx,t')
    -i\partial_{t'}
     \big) \psi(\bfx,t';t)
    \\
    +g|\psi(\bfx,t';t)|^2\psi(\bfx,t';t),
\end{multline}
with the normalization $\int\!\!d\bfx d t\;|\psi(\bfx,t';t)|^2=T$.
This normalization is a direct consequence of the fact that the physically relevant wave function is obtained as $\psi(\bfx,t)=\psi(\bfx,t,t)$.
Equation~\ref{eqn: Gross-Pitaevskii with two times} has the form of a nonlinear Schr\"odinger equation, with the linear part being static and the related wave function lives in an extended space given by the tensorial product of conventional Hilbert space and time-periodic functions, which was introduced by Sambe~\cite{SambeSteady1973}.
This implies that Eq.~\eqref{eqn: Gross-Pitaevskii with two times} can formally be derived from the Hamilton functional
\begin{equation}
 \label{eqn: Hamiltonian Gross-Pitaevskii two times}
 H_\mathrm{GP}'=
%  \frac{1}{T}
 \int\!\! d\bfx\, dt'\,
        \psi^*\big(H_0(\bfx,t')-i\partial_{t'}\big)\psi
          +\frac{g}{2}|\psi(\bfx,t')|^4.
\end{equation}
Note that, due to the integral over over time ($t'$), the functional $H_\mathrm{GP}'$ has the unit of an action instead of energy.
However, the equation of motion~\eqref{eqn: Gross-Pitaevskii with two times} can be obtained from $H_\mathrm{GP}'$ by applying Hamilton's equations.
% maybe cite some continuum mechanics book?
The minima of this Hamiltonian are stationary states of the equation of motion~\eqref{eqn: Gross-Pitaevskii with two times}.
These stationary solutions yield states, whose associated physically relevant solutions (obtained by $t'=t$) are $T$-periodic up to a global phase factor $e^{-i \varepsilon T}$.
They are the analog to Floquet states from the linear Schr\"odinger equation, and are named nonlinear Floquet states~\cite{HolthausTowards2001,Morales_Resonant_2008}. Accordingly, $\varepsilon$ is the corresponding quasienergy.

Here, the extended $(t,t')$-formalism can be used to derive a determining equation for nonlinear Floquet states.
By applying the variational principle to equation~\eqref{eqn: Hamiltonian Gross-Pitaevskii two times}, one obtains
\begin{equation}
 \label{eqn: determination nonlinear Floquet states}
 \varepsilon\psi(\bfx,t')
  =\Big(H_0(\bfx,t')-i\partial_{t'}
   +g|\psi(\bfx,t')|^2\Big)\psi(\bfx,t').
\end{equation}
This equation has been previously introduced in Ref.~\cite{Wuester_macroscopic_2012}, but was not brought into the context with the (extended) $(t,t')$-formalism.

\section{Convergence study}
\label{sec: Convergence study}
\begin{figure}[tb]
 \centering
 \includegraphics[width=0.48\textwidth]{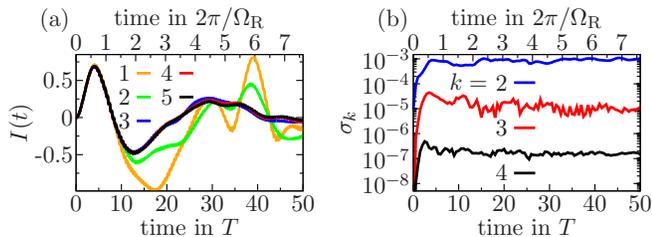}
 \caption{\textit{Convergence of MCTDHB simulations:} (a) instantaneous particle current over time for various values of $M$, given in the key; (b) occupation of modes with the same magnitude of angular momentum, defined in Eq.~\eqref{eqn: combined normalized occupation}. The system parameters are the same as in Fig.~\ref{fig: comparison MCTDHB I}b (chaotic motion).}
 \label{fig: convergence study}
\end{figure}
Within MCTDHB the maximal number of natural orbitals that are taken into account is controlled via the parameter $M$, which determines the maximal number of nonzero eigenvalues of the SPDM.
For $M=1$ the GP approximation is obtained; $M=2$ allows the description of a Bose gas occupying two natural orbitals, and so on.
Accordingly, the value of $M$ has to be chosen sufficiently high in order to obtain trustworthy results.

Fig.~\ref{fig: convergence study}a shows the time-evolution of the ratchet current for various values of $M$.
We see that the curves undergo drastic changes when $M$ is increased from $1$ to $2$, and also from $2$ to $3$.
However, when $M$ is increased further, these changes become minor.
The difference in the value of the current for $M=4$ and $M=5$ stays below $0.06$ for the considered time range.
For the results presented in Section~\ref{sec: numerical cross check}, we chose $M=4$.
However, the relative difference $\sigma(t)$ can approach values up to $200$ near those time points where the current vanishes.
Note that the relative difference between two curves diverges whenever their roots do not coincide exactly.

The influence of the single-particle basis has also been checked. For this study we have considered approximations up to $M=3$ and used different numbers of modes. We observed that angular momentum modes beyond the $k=4$ mode are only weakly occupied, which is why we dropped them for the more accurate calculations, and used a basis of nine angular momentum eigenmodes ranging from $k=-4$ to $k=4$.
Figure~\ref{fig: convergence study}b shows the combined normalized occupation of those modes with the same magnitude of angular momentum, given by
\begin{equation}
 \label{eqn: combined normalized occupation}
 \sigma_k\equiv \braket{\hat{n}_k+\hat{n}_{-k}}/N.
\end{equation}
Considered are those values of $k$ that lie outside the truncated three-level description ($k=2,\,3,\,4$).
As can be seen, the contribution of outside modes drops roughly by factors of $100$ for increasing value of $k$.
This implies that for the precision of the results presented here, the contribution of the modes $\ket{\pm4}$ is sufficiently small.
The normalized occupation of the angular momentum modes ($k=\pm4$) is below $5\times10^{-6}$ during the considered time range. This upper bound for $\sigma_4$ during the first $50$ driving cycles holds for all values of $g$.

The employed MCTDHB code uses adaptive step-size integrators which are adjusted to a high accuracy, such that the results are trustworthy. We have checked the influence of integration accuracy in the beginning of the calculations and adjusted step size tolerance to a sensible value, which was used throughout our study in the following.

\bibliographystyle{apsrev4-1}
\bibliography{bibliography}

\end{document}